\documentstyle[prl,aps,epsf]{revtex}

\begin{document}
%%\draft

%%\preprint{FERMILAB--Pub--XXX}

\title{Constraining dark energy with SNe Ia and large-scale structure}

\author{Saul Perlmutter$^{1}$, Michael S. Turner$^{2,3}$ and
Martin White$^{4}$}

\address{$^1$Institute for Nuclear and Particle Astrophysics,
E.O. Lawrence Berkeley National Laboratory,
Berkeley, CA~~94720\\
$^2$Departments of Astronomy \& Astrophysics and of Physics,
Enrico Fermi Institute, The University of Chicago,
Chicago, IL~~60637-1433\\
$^3$NASA/Fermilab Astrophysics Center,
Fermi National Accelerator Laboratory,
Batavia, IL~~60510-0500\\
$^4$Departments of Astronomy and of Physics,\\
University of Illinois at Urbana-Champaign, Urbana, IL~~61801}

\date{\today}
\maketitle

\begin{abstract}
Measurements of the distances to SNe Ia have produced strong evidence that
the expansion of the Universe is accelerating, implying the existence of a
nearly uniform component of dark energy with negative pressure.
We show that constraints to this mysterious component based upon large-scale
structure nicely complement the SN Ia data, and that together they require
$\Omega_X\in(0.6,0.7)$ and $w_X<-0.6$ (95\% cl), for
the favored flat Universe.  Other cosmological data support this conclusion.
The simplest explanation, a cosmological constant, is consistent
with this, while some of the other possibilities are not.
\end{abstract}


\section{Introduction}

Two groups \cite{scp,hi-z} have presented strong evidence that the expansion
of the Universe is speeding up, rather than slowing down.  It comes
in the form of distance measurements to some fifty supernovae of type Ia
(SNe Ia), with redshifts between 0 and 1.
The results are fully consistent with the existence of a cosmological constant
(vacuum energy)
whose contribution to the energy density is around 70\% of the critical density
($\Omega_\Lambda \sim 0.7$).
Other measurements indicate that matter alone contributes
$\Omega_M = 0.4\pm 0.1$ \cite{matter}.
Taken together, matter and vacuum energy account for an amount close to the
critical density, consistent with measurements of the anisotropy of the cosmic
microwave background (CMB) \cite{Efstathiou}.

In spite of the apparent success of the cosmological constant explanation,
other possibilities have been suggested for the ``dark energy.''
This is in part because of the checkered history of the cosmological constant:
It was advocated by Einstein to construct a static universe and discarded
after the discovery of the expansion; it was revived by Hoyle and Bondi and
Gold to solve an age crisis, later resolved by a smaller Hubble constant,
and it was put forth to explain the abundance of quasars at $z\sim 2$,
now known to be due to galactic evolution.
Further, all attempts to compute the value of the cosmological constant,
which in modern terms corresponds to the energy associated with the quantum
vacuum, have been wildly unsuccessful \cite{rmp-weinberg}.
Finally, the presence of a cosmological constant makes the present epoch
special: at earlier times matter (or radiation) dominated the energy density
and at later times vacuum energy will dominate (the ``why now?'' problem).

The key features of an alternative form for the dark energy are:
bulk pressure that is significantly negative, $w< -1/3$, where
$w\equiv p/\rho$, and the inability to clump effectively.
The first property is needed to ensure accelerated expansion and to avoid
interfering with a long matter-dominated era during which structure forms;
the second property is needed so that the dark energy escapes detection in
gravitationally bound systems such as clusters of galaxies.
Candidates for the dark energy include \cite{turnerwhite97}:
a frustrated network of topological
defects (such as strings or walls), here $w=-{n\over 3}$ ($n$ is the dimension
of the defect) \cite{frustrated} and an evolving scalar field, where
$\rho={1\over 2}\dot{\phi}^2+V(\phi)$ and $p={1\over 2}\dot{\phi}^2-V(\phi)$
(referred to by some as quintessence) \cite{rollingphi,RatraPeebles}.

The SN Ia data alone do not yet discriminate well against these different
possibilities \cite{scp,w-constraints}.  As shown in Fig.~\ref{fig:composite},
the maximum likelihood region in the $\Omega_M$--$w$ plane runs roughly
diagonally: less negative pressure is permitted if the fraction of critical
density contributed by dark energy is larger.
Following earlier work \cite{turnerwhite97}, this led us to consider other
cosmological constraints: large-scale structure, anisotropy of the CMB, the
age of the Universe, gravitational lensing, and measurements of the Hubble
constant and of the matter density.
As we shall show, some of the additional constraints, especially large-scale
structure, complement the SN Ia constraint, and serve to sharpen the limits
to $\Omega_M$ and $w$; others primarily illustrate the consistency of these
measurements with the SN Ia result.
In the end, we find $\Omega_X\in(0.6,0.7)$ and $w<-0.6$ (95\% cl).

\section{Method}

Our underlying cosmological paradigm is a flat, cold dark matter model with
a dark-energy component, though as we will discuss later our
results are more general.  We restrict ourselves to flat models both because
they are preferred by the CMB anisotropy data and a flat
Universe is strongly favored by inflation.
We restrict ourselves to cold dark matter models because of the success of
the cold dark matter paradigm and the lack of a viable alternative.
For our space of models
we construct marginalized likelihood functions based upon
SNe Ia, large-scale structure, and other cosmological measurements,
as described below.

Our model parameter space includes the usual cosmological parameters
($\Omega_M$, $\Omega_{B}h^2$, and $h$) and the amplitude and spectral
index of the
spectrum of Gaussian curvature fluctuations ($\sigma_8$ and $n$).
For the dark-energy component, we choose to focus on the dynamical
scalar-field models, because the frustrated defect models are
at best marginally consistent with the SN Ia data alone
\cite{scp,w-constraints}.

In the dynamical scalar-field models the
equation of state $w\equiv p/\rho$ varies with time.  However for
most of our purposes, only one additional free parameter needs
to be specified, an
``effective'' equation of state.  We choose $\widehat{w}_{\rm eff}$ to be
that value $w$ which, if the Universe had $w$ constant, would reproduce the
conformal age today.  We choose this definition because
the CMB anisotropy spectrum and the {\sl COBE\/} normalization of the matter
power spectrum remain constant (to within 5--10\%) for different scalar field
models with the same $\widehat{w}_{\rm eff}$ \cite{White98}.
For the models under consideration $\widehat{w}_{\rm eff}$ is closely
approximated by \cite{HWDCS}
\begin{equation}
  w_{\rm eff} \equiv \int da\ \Omega_\phi(a) w(a)/\int da\ \Omega_\phi(a) \,.
\label{eqn:weff}
\end{equation}
and, since it is simpler to compute, we have used $w_{\rm eff}$ throughout.
Obviously, our results also apply to constant $w$ models (e.g., frustrated
defects), by taking $w=w_{\rm eff}$.

While $w_{\rm eff}$ neatly parameterizes the scalar-field models from the
standpoint of large-scale structure and the CMB anisotropy,
it does not do as well when it comes to the SN Ia data.  Recall that
$w_{\rm eff}$ as defined in Eq.~(\ref{eqn:weff}) receives a contribution
from a wide range of redshifts.  The SN Ia data however are sensitive mostly
to $z\sim 1/2$.  Since $w$ becomes less negative with time in the models we
are considering, the SN Ia data ``see'' a less negative $w$ than the CMB by
a model dependent amount.  We shall return to this point later.

We normalize our models to the {\sl COBE\/} 4-year data \cite{cobe4yr}
using the method of Ref.~\cite{BunWhi}.  Beyond the {\sl COBE\/}
measurements, the small-scale
anisotropy of the microwave background tells us that the Universe is close
to being spatially flat (position of the first acoustic peak) and that
$\Omega_M$ is less than one and/or the baryon density is high
(height of the first acoustic peak).
We have not included a detailed fit to the current data
(see e.g.~Ref.~\cite{Efstathiou}), but rather impose flatness.
The additional facts that might be gleaned from present CMB measurements,
$\Omega_M < 1$ and high baryon density, are in fact much more strongly
imposed by the large-scale structure data and the Burles -- Tytler
deuterium measurement.

We require that the power-spectrum shape fit the redshift-survey
data as compiled by Ref.~\cite{shape} (excluding the 4 smallest scale points
which are most sensitive to the effects of bias and nonlinear effects).
On smaller scales we require that all of our models reproduce the observed
abundance of rich clusters of galaxies.
This is accomplished by requiring $\sigma_8 =  \left( 0.55\pm 0.1 \right)
\Omega_M^{-0.5}$, where $\sigma_8$ is the rms mass fluctuation in spheres
of $8\,h^{-1}\,$Mpc computed in linear theory \cite{sigma8-refs}.
The baryon density is fixed at the central value indicated by the
Burles--Tytler deuterium measurements, $\Omega_Bh^2 = 0.019\pm 0.001$
\cite{burlestytler98}.
We assume that clusters are a fair sample of the matter
in the Universe so that the cluster baryon fraction $f_B =(0.07\pm 0.007)
h^{-3/2}$ reflects the universal ratio of baryons to
matter ($\Omega_B/\Omega_M$).
We marginalize over the spectral index and Hubble constant, assuming Gaussian
priors with $n=0.95\pm 0.05$, which encompasses most inflationary models,
and $h=0.65\pm 0.05$, which is consistent with current measurements.

There are three other cosmological constraints that
we did not impose: the age of the
Universe, $t_0=(14\pm 2)$Gyr \cite{Chaboyeretal}; direct measurements
of the matter density, $\Omega_M=0.4\pm 0.1$, and the frequency
of multiply imaged quasars.
While important, these constraints serve to prove consistency, rather than
to provide complementary information.
For example, the SN Ia data together with our Hubble constant constraint
lead to an almost identical age constraint \cite{scp,hi-z}.  
The lensing constraint, recently studied in detail for dynamical scalar-field
models \cite{waga}, excludes the region of large $\Omega_X$ and very negative
$w$ (at 95\% cl, below the line $w_{\rm eff} = -0.55 - 1.8\Omega_M$),
which is disfavored by the SN Ia data.  The matter density
determined by direct measurements, $\Omega_M=0.4\pm 0.1$, is
consistent with that imposed by the LSS and Hubble constant constraints.

\section{Results}

\begin{figure}
\centerline{\epsfxsize=15cm \epsfbox{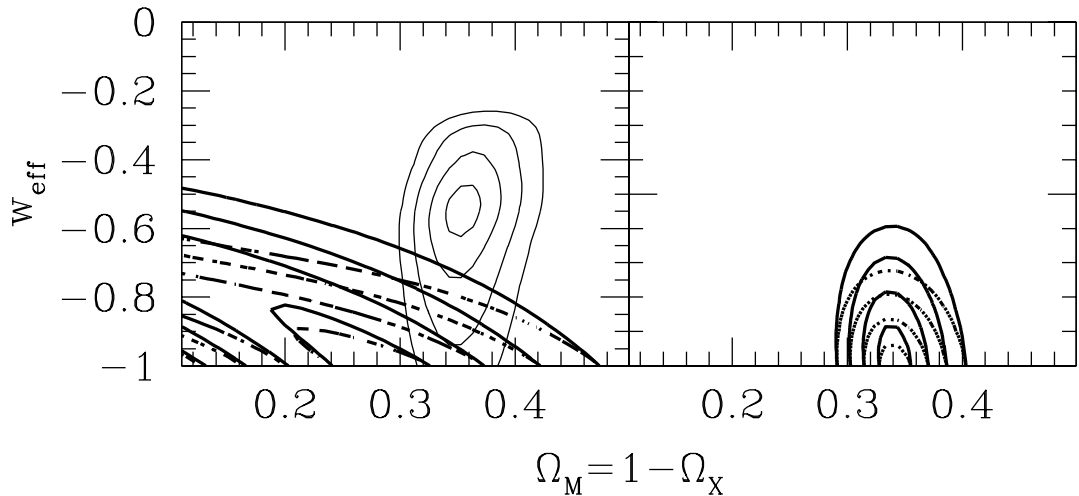}}
\caption{Contours of likelihood, from $0.5\sigma$ to $2\sigma$, in the
$\Omega_M$--$w_{\rm eff}$ plane.  Left:  The thin solid lines are the
constraints from LSS and the CMB.  The heavy lines are the SN Ia
constraints (using the Fit C supernovae of Ref.~\protect\cite{scp}) for
constant $w$ models (solid curves) and for a scalar-field model
with an exponential potential (broken curves; quadratic and quartic
potentials have very similar SN Ia constraints).
Note that the SN Ia contours for dynamical scalar-field
models and constant $w$ models are slightly offset (see text).
Right:  The likelihood contours from all of our
cosmological constraints for constant $w$ models (solid)
and dynamical scalar-field models (broken).}
\label{fig:composite}
\end{figure}

As can be seen in Fig.~\ref{fig:composite}, our large-scale structure and
CMB constraints neatly complement the SN Ia data.
LSS tightly constrains $\Omega_M$, but is less restrictive along
the $w_{\rm eff}$ axis.
This is easy to understand: in order to fit the power spectrum data, a
{\sl COBE\/}-normalized CDM model must have ``shape parameter''
$\Gamma = \Omega_M h \sim 0.25$ (with a slight dependence on $n$).
Together with the constraint $h=0.65\pm0.05$ (and our $f_B$
constraint) this leads to $\Omega_M \sim 0.35$.
As discussed in Ref.~\cite{turnerwhite97}, the $\sigma_8$ constraint can
discriminate against $w_{\rm eff}$; however, allowing the spectral index to
differ significantly from unity diminishes its power to do so.

Note that the SN Ia likelihood contours for the dynamical scalar-field model
and the constant-$w$ models {\em are not\/} the same while the LSS contours
are identical.  With the Fit C supernovae of Ref.~\cite{scp} and the dynamical
scalar-field models considered here (quadratic, quartic and exponential
scalar potentials), the contours are displaced by about 0.1 in $w_{\rm eff}$:
the 95\% cl upper limit to $w_{\rm eff}$ for the constant $w$ models is
$-0.62$, while for the quartic, quadratic and exponential potentials for
$V(\phi )$ it is $-0.75$, $-0.76$ and $-0.73$ respectively.
The reason for this shift is simple: the $w$ dependence of LSS is almost
completely contained in the distance to the last-scattering surface and
$w_{\rm eff}$ is constructed to hold that constant.
On the other hand, the $w$ dependence of the SN Ia results is more heavily
weighted by the recent value of $w$; said another way, there is a different
effective $w$ for the SN Ia data.
{\em This fact could ultimately prove to be very important in discriminating
between different models.}

Additionally there are a class of dynamical scalar-field models that have
attracted much interest recently \cite{RatraPeebles,ZWS}.
For these potentials (here we consider $V(\phi ) = c/\phi^p$ and
$V(\phi)=c[e^{1/\phi}-1]$), and a wide range of initial conditions the
scalar-field settles into a ``tracking solution'' that depends only upon
one parameter (here $c$) and the evolution of the cosmic scale factor,
suggesting that they might help to address the ``why now?'' problem.

\begin{figure}
\centerline{\epsfxsize=15cm \epsfbox{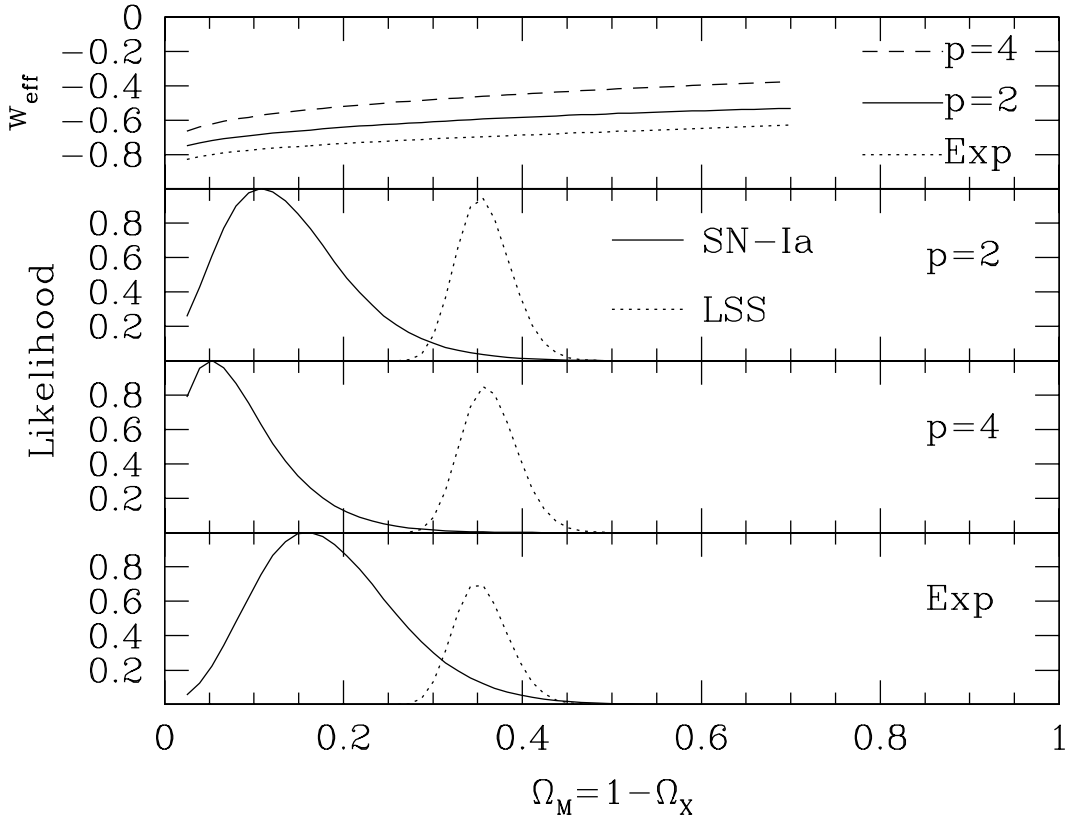}}
\caption{Upper panel: The relationship between $w_{\rm eff}$ and $\Omega_M$
for a selection of tracker potentials.  Lower panels: the CMB and LSS
likelihoods from Fig.~\protect\ref{fig:composite} as
a function of $\Omega_M$ (dotted) and the SN Ia likelihood (solid --
normalized to unity at the peak).
As can be seen clearly, tracker models have difficulty simultaneously
accommodating the SN Ia and LSS constraints.}
\label{fig:tracker}
\end{figure}

For our purposes, the most interesting fact is that each tracker potential
picks out a curve in $\Omega_M-w_{\rm eff}$ space.
Typically the lower values of $\Omega_M$ go with the most negative values
of $w_{\rm eff}$ and vice versa (see Fig.~\ref{fig:tracker})
This fact puts the tracker solutions in jeopardy, as shown in the same figure.
For the tracker models shown here ($p=2,4$ and exponential), the 95\% cl
intervals for the SN Ia and LSS data barely overlap.  The situation is
even worse for larger values of $p$.  A similar problem was noted in
Ref.~\cite{SWZ}.

Finally, we comment on the robustness of our results.  While we have
restricted ourselves to flat models, as preferred by the CMB data,
our constraints do not depend strongly on this assumption.  This is
because the LSS constraints are insensitive to the flatness assumption,
and curvature, which corresponds to a $w_{\rm eff}=-{1\over 3}$ component,
is strongly disfavored by the SN Ia results.  We have not explicitly
allowed for the possibility that inflation-produced gravity waves account for
a significant part of the CMB anisotropy on large-angular scales
(i.e., $T/S > 0.1$), which would have the effect of decreasing the overall
amplitude of the {\sl COBE\/} normalized power spectrum.
In fact, allowing for gravity waves would not change our results, as this
degree of freedom is implicitly accounted for by a combination of $n$, the
normalization freedom in the power spectrum and the uncertainty in the
{\sl COBE\/} normalization.  

Our model space does not explore more radical possibilities, for example,
that neutrinos contribute significantly to the mass density
or a nonpower-law or isocurvature spectrum
of density perturbations \cite{Peebles98}.
Even allowing for these possibilities (or others) would not change our
results significantly if one still adopted the mass density constraint,
$\Omega_M =0.4\pm 0.1$.  As discussed earlier,
it is almost as powerful as the CDM-based LSS constraint.

\section{Conclusions}

The evidence provided by SNe Ia that the Universe is accelerating rather than
slowing solves one mystery -- the discrepancy between direct measurements of
the matter density and measurements of the spatial curvature based upon CMB
anisotropy -- and introduces another -- the nature of the dark energy that
together with matter accounts for the critical density.  SNe Ia alone do
not yet strongly constrain the nature of the dark energy.

In this {\em Letter\/} we have shown that consideration of other important
cosmological data both complement and reinforce the SN Ia results.
In particular, as illustrated in Fig.~\ref{fig:composite}, consideration of
large-scale structure leads to a constraint that nicely complements the SN Ia
constraint and strengthens the conclusions that one can draw.
Other cosmological constraints  --  age of the Universe, frequency of
gravitational lensing and direct measures of the matter density -- provide
information that is consistent with the SN Ia constraint (lensing and age)
and the LSS constraint (matter density), and thereby reinforces the self
consistency of the whole picture of a flat Universe with cold dark matter
and dark energy.

Finally, what have we learned about the properties of the dark-energy
component?  The suite of cosmological constraints that we have applied
indicate that $\Omega_X\in (0.6,0.7)$ and $w_{\rm eff}<-0.6$ (95\% cl),
with the most likely value of $w_{\rm eff}$ close to $-1$
(see Fig.~\ref{fig:composite}).
The frustrated network of light cosmic string ($w_{\rm eff}=-{1\over 3}$)
is strongly disfavored, and a network of frustrated walls
($w_{\rm eff}=-{2\over 3}$) is only slightly more acceptable.
Also in the disfavored category are tracker models with $V(\phi )=c/\phi^p$
and $p=2,4,6,8,\cdots$.
Dynamical scalar-field models can be made acceptable provided $w_{\rm eff}$
is tuned to be more negative than $-0.7$.
The current data definitely prefer the most economical, if not the most
perplexing, solution: Einstein's cosmological constant.

\acknowledgments
This work was supported by the DoE (at Chicago, Fermilab, and Lawrence
Berkeley National Laboratory) and by the NASA (at Fermilab by grant
NAG 5-7092).  MW is supported by the NSF.


\begin{thebibliography}{99}

\bibitem{scp}
S. Perlmutter et al, LBL-42230 (1998)(astro-ph/9812473);
S. Perlmutter et al, Astrophys. J., in press (1999) (astro-ph/9812133).

\bibitem{hi-z}
B. Schmidt et al, Astrophys. J. {\bf 507}, 46 (1998).
A.G. Riess, et al., Astron. J., in press (astro-ph/9805200)

\bibitem{matter}
See e.g., A. Dekel et al, in {\it Critical Dialogues
in Cosmology}, ed. N. Turok (World Scientific, Singapore, 1997);
and M.S. Turner, Pub. Astron. Soc. Pac., in press (1999)
(astro-ph/9811454) and references therein.

\bibitem{Efstathiou}
G. Efstathiou et al, astro-ph/9812226.

\bibitem{rmp-weinberg}
S. Weinberg, Rev. Mod. Phys. {\bf 61}, 1 (1989).

\bibitem{turnerwhite97}
M.S. Turner and M. White, Phys. Rev. D {\bf 56}, R4439 (1997).

\bibitem{frustrated}
A. Vilenkin, Phys. Rev. Lett. {\bf 53}, 1016 (1984);
D. Spergel and U.-L. Pen,  Astrophys. J. {\bf 491}, L67 (1997).

\bibitem{rollingphi}
M. Bronstein, Phys. Zeit. Sowjet Union {\bf 3}, 73 (1933);
M. Ozer and M.O. Taha, Nucl. Phys. B {\bf 287} 776 (1987);
K. Freese et al., {\it ibid} {\bf 287} 797 (1987);
L.F. Bloomfield-Torres and I. Waga, Mon. Not. R. astron. Soc.
{\bf 279}, 712 (1996); J. Frieman et al, Phys. Rev. Lett.
{\bf 75}, 2077 (1995); K. Coble et al, Phys. Rev. D {\bf 55},
1851 (1996); R. Caldwell et al, Phys. Rev. Lett. {\bf 80},
1582 (1998).

\bibitem{RatraPeebles}
B. Ratra and P.J.E. Peebles, Phys. Rev. D {\bf 37}, 3406 (1988).

\bibitem{w-constraints}
P. Garnavich et al, Astrophys. J. {\bf 509}, 74 (1998).


\bibitem{White98}
M. White, Astrophys. J. {\bf 506}, 495 (1998).

\bibitem{HWDCS}
G. Huey et al, astro-ph/9804285.

\bibitem{cobe4yr}
C.L. Bennett et al., Astrophys. J. {\bf 454}, L1 (1996).

\bibitem{BunWhi}
E. Bunn and M. White, Astrophys. J. {\bf 480}, 6 (1997).

\bibitem{shape}
J. Peacock and S. Dodds, Mon. Not. R. astron. Soc. {\bf 267}, 1020 (1994).

\bibitem{sigma8-refs}
See e.g., S.D.M. White, G. Efstathiou, and C.S. Frenk,
Mon. Not. R. astron. Soc. {\bf 262}, 1023 (1993);
V.R. Eke, S. Cole, C.S. Frenk, and P. Henry, {\it ibid},
{\bf 298}, 1145 (1998);
P.T.P. Viana and A.R. Liddle, {\it ibid}, in press (1999) (astro-ph/9803244).

\bibitem{burlestytler98}
S. Burles and D. Tytler, Astrophys. J.
{\bf 499}, 699 (1998); {\it ibid} {\bf 507}, 732 (1998).

\bibitem{Chaboyeretal}
B. Chaboyer et al, Astrophys. J. {\bf 494}, 96 (1998).

\bibitem{waga}
I. Waga and A.P.M.R. Miceli, astro-ph/9811460.

\bibitem{ZWS}
I.~Zlatev, L.~Wang, and P.J.~Steinhardt, astro-ph/9807002

\bibitem{SWZ}
P.J.~Steinhardt, L.~Wang, and I.~Zlatev, astro-ph/9812313

\bibitem{Peebles98}
P.J.E. Peebles, astro-ph/9805194 and 9805212.

\end{thebibliography}
\end{document}